\documentclass[aps,prl,twocolumn,psfig,showpacs]{revtex4-1}
\usepackage{amsfonts}
\usepackage{amsmath}
\usepackage{graphicx}
\usepackage{bm}
\usepackage{color}
\usepackage{amssymb}
\usepackage{ulem}
\usepackage{times}
\usepackage{dcolumn}
\usepackage{cases}
\usepackage{txfonts}

\begin{document}

\title{Honeycomb Heisenberg Spin Ladder: Unusual Ground State and Thermodynamic Properties}

\author{Yang Zhao$^{1}$, Wei Li$^{1}$, Bin Xi$^{1}$, Shi-Ju Ran$^{1}$, Yuan-Yuan Zhu$^{2}$, Bing-Wu Wang$^{2}$, Song Gao$^{2}$, Gang Su$^{1}$}
\email[Corresponding author. ]{Email:
gsu@ucas.ac.cn}
 \affiliation{$^{1}$Theoretical Condensed Matter Physics and Computational
Materials Physics Laboratory, School of Physics, University of Chinese Academy
of Sciences, P. O. Box 4588, Beijing 100049, China
\linebreak $^{2}$College of Chemistry and Molecular Engineering, State Key Laboratory of Rare Earth Materials Chemistry and Applications, Peking University, Beijing 100871, China}

\begin{abstract}
The unusual ground state and thermodynamic properties of spin-1/2 two-leg honeycomb (HC) spin ladder are systematically studied by jointly utilizing various analytical and numerical methods. The HC spin ladder is found to exhibit behaviors dramatically different from those of the conventional square spin ladder. A strong relevant term $\vec{n}\cdot\vec{J}$ and a half saturation magnetization plateau that can be attributed to the formation of diluted dimer states are observed in the HC ladder, both of which are absent in the square ladder. The ground state phase diagram of the HC spin ladder is identified, and the thermodynamic properties of the specific heat and susceptibility for different couplings are thoroughly explored, where two kinds of excitations are unveiled. The distinct Wilson ratios for both spin ladders at the lower critical fields are also obtained. Our calculated result is well fitted to the experimental data of the two-leg HC spin ladder compound $[Cu_2L^1(N_3)_4]_n$.
\end{abstract}

\maketitle

Quantum spin ladders have attracted intense attention from both experimental and theoretical aspects for their rich physical properties (e.g. Ref. [\onlinecite{TMRice}]) as well as diverse applications (for instance in superconductivity \cite{Tranquada}, field-induced quantum phase transitions \cite{Kohama}, quantum computation \cite{Suncp}, high energy physics \cite{Lake}, and so on). For conventional square spin ladders with all antiferromagnetic (AF) interactions, it has been shown that those with even number of legs have a finite spin gap and exhibit magnetic short-range orders, while those with odd number of legs display gapless magnetic excitations, revealing that the crossover from spin ladders to a two-dimensional antiferromagnet is not smooth \cite{TMRice}. In magnetic fields, spin ladders could close the energy gaps and show the Tomonaga-Luttinger liquid (TLL) behavior with the linear temperature dependence of specific heat at low temperature. In recent years, the exotic properties of intriguing spin ladder materials like the S=1/2 two-leg compounds with AF legs $(C_5H_{12}N)_2CuBr_4$ (BPCB) \cite{Ruegg} and $(C_7H_{10}N)_2CuBr_4$ (DIMPY) \cite{Hong}, with ferromagnetic (F) legs 3-Cl-4-F-V [3-(3-chloro-4-fluorophenyl)-1,5-diphenylverdazyl] \cite{Yamaguchi}, and with frustrations along the legs $BiCu_2PO_6$ \cite{Tsirlin}, etc., have been extensively explored, where novel quantum states and unconventional spinon excitations were disclosed \cite{Yamaguchi,TLLW,Schmidiger,Choi,Casola}, illustrating that the spin ladders continue to surprise us in the area of strongly correlated quantum systems.

\begin{figure}[tbp]
  \centering
  \includegraphics[width=0.9\linewidth]{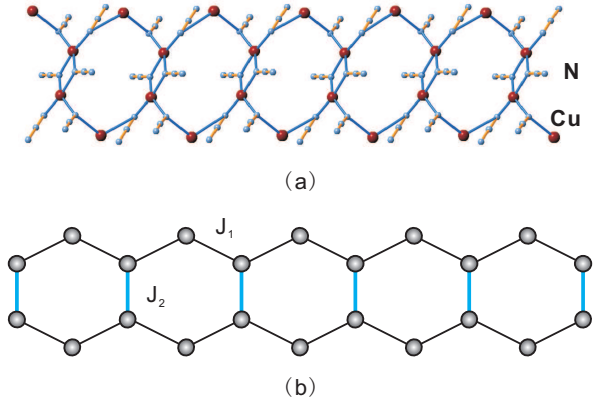}\\
  \caption{(Color online) (a) Crystal structure of the azide copper coordination compound $[Cu_2L^1(N_3)_4]_n$, where Cu ions have spin 1/2 and form a honeycomb spin ladder \cite{hcladder}. (b) The two-leg honeycomb spin ladder, where J$_1$ and J$_2$ stand for the exchange coupling along the leg and the rung, respectively.}\label{Structure}
\end{figure}

Quite recently, a two-leg honeycomb (HC) spin ladder system has been realized in a four azide copper coordination compound $[Cu_2L^1(N_3)_4]_n$ (L$^1$=2, 6-bis (4, 5-dihydrooxazol-2-yl) pyridine) \cite{hcladder}, whose crystal structure is shown in Fig. \ref{Structure} (a). In this material, copper ions with spin $\frac{1}{2}$ are coupled by $N_3$ bridges, and form a two-leg HC spin ladder. The
experimental analysis demonstrates that the single $N_3$ bridges give rise to an intra-chain AF coupling $J_1$, and the double $N_3$ bridges lead to an inter-chain F coupling $J_2$, as indicated in Fig. \ref{Structure} (b). Motivated by this compound, an interesting question arises naturally: Does there exist a fundamental difference of physical properties between two-leg conventional square and unconventional HC spin ladders? In this Letter, by jointly using various methods including bosonization, quantum Monte Carlo (QMC), the infinite time evolution block decimation (iTEBD) \cite{Vidal} and the linearized tensor renormalization group (LTRG) \cite{LiW}, we shall answer this question by comparatively studying the ground state and thermodynamic properties of these two kinds of spin ladders.

Let us begin with the Hamiltonian of the S=1/2 two-leg HC spin ladder given by
\begin{eqnarray}
H&=&J_1\sum_{l=1,2}\sum_i\mathbf{S}_{i,l}\cdot\mathbf{S}_{i+1,l}
+J_2\sum_i\mathbf{S}_{2i-1,1}\cdot\mathbf{S}_{2i-1,2} \nonumber \\
  &-&g\mu_Bh\sum_{l=1,2}\sum_i\mathbf{S}^{z}_{i,l}, \label{hamil}
\end{eqnarray}
where $\mathbf{S}_{i,l}$ denotes the spin-$\frac{1}{2}$ operator on site $i$ of the $l$th leg ($l=1, 2$), $J_1$ and $J_2$ are exchange interactions along the leg and the rung, respectively, $g$ is the Land\'e factor, $\mu_B$ is the Bohr magneton, and $h$ is the magnetic field. For simplicity, we take $g\mu_B=1$, $k_B=1$ and $J_1$ as energy scale henceforth. In the following, three interesting cases will be considered: (i) $J_1$, $J_2>0$; (ii) $J_1<0$, $J_2>0$; (iii) $J_1>0$, $J_2<0$. We shall invoke the QMC to calculate the spin gaps of both ladders, where the length of the leg is assumed $L=1000$ and the error is kept less than 10$^{-5}$. The iTEBD is employed to study the magnetic curves of the HC ladder in the ground state, where the bond dimension is taken as $D_c=500$ and the truncation error is kept less than 10$^{-7}$. The LTRG method, which is quite accurate and efficient especially at very low temperature (e.g. Ref. [\onlinecite{yanxin}]), is applied to investigate the thermodynamic properties of both kind of ladders, where $D_c=800$, the Trotter step is $\tau=0.1$, and the truncation error is kept less than 10$^{-9}$.

It is known that the two-leg square spin ladders with all AF couplings always have gapful excitations in the absence of a magnetic field. When the inter-chain coupling is ferromagnetic (J$_2<0$), the spin gap is of Haldane-like, as the two 1/2 spins on the rung form a spin triplet; when J$_2$ is antiferromagnetic (J$_2>0$), the spin gap is brought by strong relevant perturbations from the inter-chain couplings, which may lead to the singlet-triplet gap as in the large J$_2$ limit. For the two-leg HC spin ladder, there are only half number of rungs as many as those of the conventional square spin ladder. One may therefore anticipate that it owns different behaviors from the square spin ladder. By using the bosonization method \cite{typical gap}, the effective model of the two-leg HC spin ladder for J$_2 \ll$ J$_1$ can be obtained:
\begin{eqnarray}
  \mathcal{H} &=& \mathcal{H}_0+\mathcal{H}^{'}, \\
  \mathcal{H}_0 &=& \sum_{l=1}^{2}\frac{2\pi v_s}{3}(:\vec{J}_{l,R}\cdot\vec{J}_{l,R}:+:\vec{J}_{l,L}\cdot\vec{J}_{l,L}:), \\
  \mathcal{H}^{'} &=& J_2(\vec{J}_{1}\cdot\vec{J}_{2}+\vec{J}_{1}\cdot\vec{n}_{2}+\vec{n}_{1}\cdot\vec{J}_{2}+\vec{n}_{1}\cdot\vec{n}_{2}),
\end{eqnarray}
where $\mathcal{H}_0$ is the free part and $\mathcal{H}^{'}$ is the perturbation term. In above equations, $\vec{J}_{l, L (R)}$ is the left (right) moving SU(2) current operator that represents the smooth magnetization part of spin density operators on the $l$th leg, $v_{s}\sim J_{1}a_{0}$ is the spin velocity with $a_{0}$ the lattice constant, $\vec{J}_{l}=\vec{J}_{l,L}+\vec{J}_{l,R}$, and $\vec{n}_{l}$  is the staggered part of the spin density operator
on the $l$th leg whose scaling dimension is $\frac{1}{2}$. For the conventional square spin ladder, the two terms $\vec{J}_{1}\cdot\vec{n}_{2}$ and $\vec{n}_{1}\cdot\vec{J}_{2}$ in Eq. (4) are absent and for small $J_2/J_1$, the spin gap opens due to the strong relevant perturbation $\vec{n}_1\cdot\vec{n}_2$ induced by the rung coupling \cite{typical gap}. In contrast, for the HC spin ladder we find that there exists the strong relevant operator $\vec{n}\cdot\vec{J}$ whose scaling dimension is $\frac{3}{2}$ that is bigger than the value of $\vec{n}_1\cdot\vec{n}_2$ in the Hamiltonian density. In accordance with the renormalization theory \cite{Tsvbook}, the increasing speed of spin gap $\Delta$ with $J_2/J_1$ in the HC ladder should be slower than that in the square spin ladder. It is interesting to mention that the appearance of the term $\vec{n}\cdot\vec{J}$ is unusual, which is sparse in other low-dimensional quantum correlated systems, and therefore makes the two-leg HC spin ladder very interesting and nontrivial.

\begin{figure}[tbp]
  \includegraphics[width=0.85\linewidth]{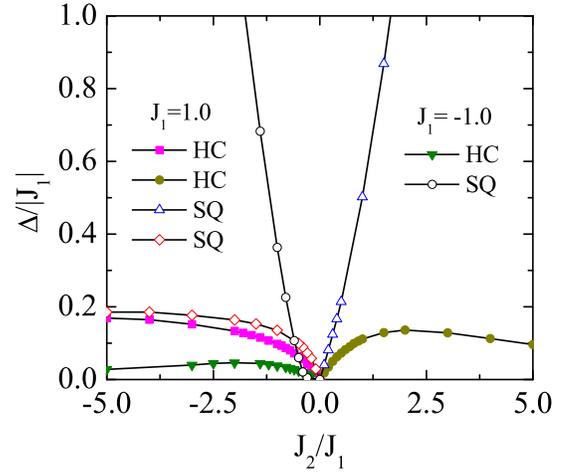}
  \caption{(Color online) Spin gap $\Delta/|J_1|$ versus $J_2/J_1$ of the spin-1/2 two-leg honeycomb (HC) and square (SQ) spin ladders. The results were obtained by the QMC calculations.}\label{gap}
\end{figure}

To substantiate the above analysis, we calculated the spin gap $\Delta/|J_1|$ as a function of $J_2/J_1$ in the two-leg square and HC spin ladders using the QMC method, respectively, where the results for the three cases are presented in Fig. \ref{gap}. It can be seen that the behaviors of the spin gap $\Delta/|J_1|$ against the coupling ratio  $J_2/J_1$ are quite different for the square and HC spin ladders, where the open symbols are adopted for the square spin ladder while the filled symbols are for the HC ladder. For case (i) and (ii) with $J_2>0$, $\Delta/|J_1|$ as a function of $J_2/J_1$ exhibits a broad hump for the HC ladder, while it increases or decreases dramatically with increasing  $J_2/J_1$ for the square ladder where $\Delta/|J_1| \propto J_2/|J_1|$ when $J_2$ is dominant regardless of the sign of $J_1$. This is understandable for the square ladder, because $\Delta$ equals the energy that is needed to break one singlet on the rung from the ground state to form a triplet, leaving $\Delta/|J_1| \propto J_2/|J_1|$. For the HC ladder, when $J_2/|J_1|$ becomes large, the spins on the rung could form spin singlets, while the other spins on the leg are nearly "free". Therefore, there should exist two kinds of excitations in the system under interest: the singlet-triplet excitation, and the single spin excitation. Using the perturbative method \cite{dispersion}, we uncovered that the two excitations are mixed and the single spin excitation lowers the excitation energy conspicuously, giving rise to a smaller but finite gap and forming a broad hump. For case (iii) with $J_2<0$, the spin gap of both ladders displays a similar behavior although the gap of the HC ladder is slightly smaller than that of the square ladder. As the spin triplets can be formed along the rung in this case, the spin gaps of both ladders show the characters of Haldane-like gap, indicating that the HC ladder in this situation belongs to the Haldane phase.

\begin{figure}
    \includegraphics[width=0.65\linewidth]{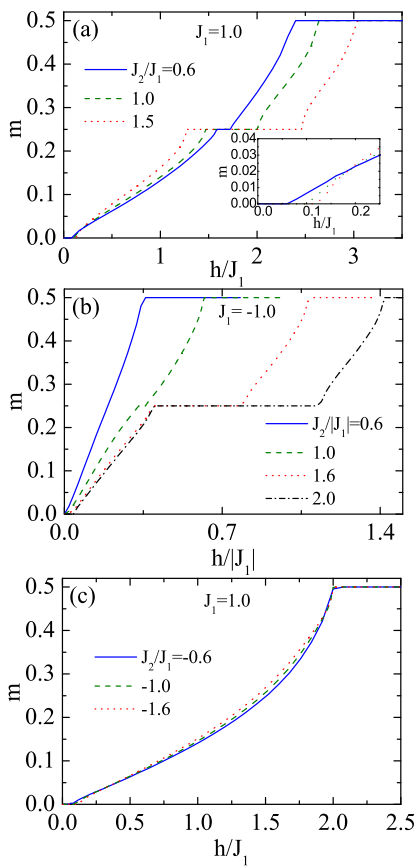}\\
  \caption{(Color online) The magnetization curves of the S=1/2 two-leg honeycomb spin ladder for three cases. (a) $J_1>0$ and $J_2>0$; (b) $J_1<0$ and $J_2>0$; (c) $J_1>0$ and $J_2<0$. The inset in (a) is the enlarged part in small magnetic fields.}
  \label{AF_AF}
\end{figure}

The magnetization curve of the HC spin ladder is quite distinct from that of the square spin ladder. In the latter case there is no magnetization plateau other than zero and fully polarized plateaux. For the HC spin ladder with AF rung coupling ($J_2>0$), besides the zero and saturation plateaux, there exists a half saturated magnetization plateau at which the magnetization per site $m=\frac{1}{4}$, as shown in Figs. \ref{AF_AF}(a) and (b). It should be noted that this result is consistent with the condition \cite{oya} for the appearance of magnetization plateau $n(S-m)=integer$, as in the present case  $S=1/2$ and the period $n=4$ that can be readily seen from the mapping of the HC ladder to an effective spin chain. In addition, the width of plateau is proportional to $J_2/|J_1|$. For $J_2<0$, there are two plateaux with $m=0$ and $m=1/2$, and the half saturated magnetization plateau does not exist [Fig. \ref{AF_AF}(c)]. This observation resembles the zigzag spin chain and rung alternating spin ladder \cite{rung_alt,di_dimer}, where the half saturated plateau was also observed. The reason behind this fact is that there exists a diluted dimer state in which spin singlet pairs and polarized spins are arranged alternatively. To confirm this, we calculated the magnetization per site at the sites on the rung and on the leg, respectively, for the HC spin ladder, and found that after closure of the $m=0$ plateau, with increasing the field the magnetization of single spins on the leg increases very fast to the nearly saturated state, while the spins on the rungs in singlet states contribute little to the total magnetization. In all three cases, we observed that the plateau with $m=0$ always exists for various $J_2/J_1$, showing the existence of the spin gap in the absence of a magnetic field, which is in consistent with the results presented in Fig. 2.

\begin{figure}
    \includegraphics[width=1.0\linewidth]{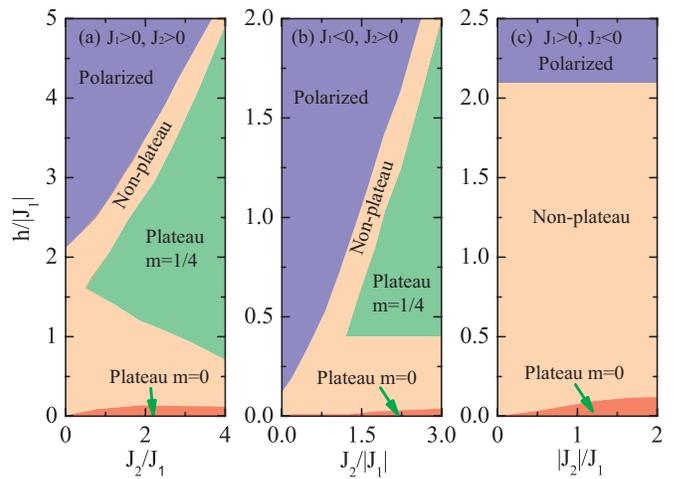}
  \caption{(Color online) Phase diagram in the plane of field ($h/|J_1|$)-coupling ratio ($J_2/J_1$) for the spin-1/2 two-leg honeycomb spin ladder in the ground state. (a) $J_1>0, J_2>0$; (b) $J_1<0, J_2>0$; (c) $J_1>0, J_2<0$. The regions where $m=0, 1/4, 1/2$ plateaux appear are separated by the non-plateau region.}
  \label{phase-diagram}
\end{figure}

By collecting the critical magnetic fields at which the magnetization plateau disappears or appears at zero temperature, we obtained the phase diagram in the plane of magnetic field-coupling ratio of the two-leg HC spin ladder in the ground state for three cases, respectively, as presented in Fig.  \ref{phase-diagram}. One may observe that the region for the appearance of $m=0$ plateau in the two cases with $J_2>0$ is much smaller than that of $m=1/4$ plateau, which reveals that the corresponding spin gap can be easily closed by a small field. The regions between those of two plateaux are of gapless excitations in the magnetic field [Figs. \ref{phase-diagram}(a) and (b)]. For the case with $J_2<0$, as there is no $m=1/4$ plateau, the gapless (non-plateau) region is much wider than that of $m=0$ plateau, as shown in  Fig. \ref{phase-diagram}(c). Thus, the three cases have distinct behaviors in the ground state, which also deviate from those of the corresponding square spin ladder.

\begin{figure}
    \includegraphics[width=1.0\linewidth]{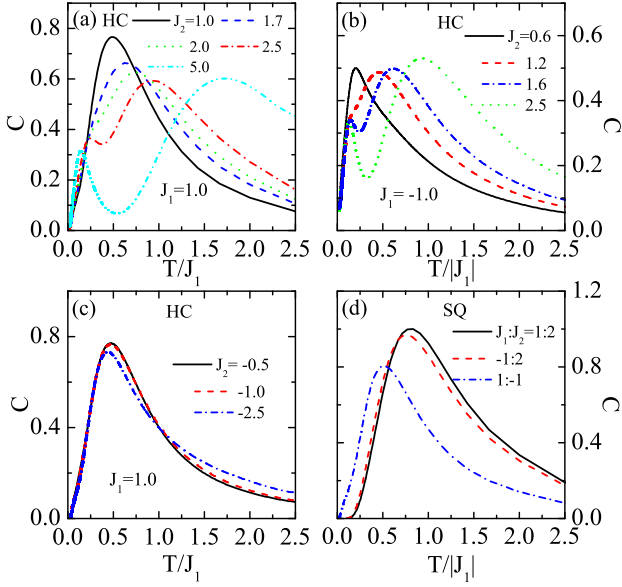}
  \caption{(Color online) The temperature dependence of specific heat $C$ in the S=1/2 two-leg spin ladders for different coupling ratios. (a)-(c): honeycomb (HC) ladder;  (d) square (SQ) ladder. The results were calculated by the LTRG method, where $D_c$=800 states are retained.}
  \label{capacity}
\end{figure}

Figure \ref{capacity} gives the temperature ($T$) dependence of the specific heat ($C$) of the two-leg HC spin ladder for various couplings. In three cases, the specific heat as a function of temperature is diverse. For $J_1, J_2>0$, when $J_2/J_1<2$, there is only one peak, where the position of the peak moves to the high temperature side with increasing $J_2/J_1$; when $J_2/J_1>2$, the peak splits into two peaks in which the left peak moves to the low temperature side while the right moves to the opposite direction, as shown in Fig.  \ref{capacity}(a). For $J_1<0$ and $J_2>0$, one may see that the specific heat shows the similar behavior as in case (i) except that the single peak splits into double peaks at a different position (about $J_2/J_1=1.2$) [Fig.  \ref{capacity}(b)]. Such a double peak behavior of the specific heat can be attributed to the two kinds of excitations \cite{two_lev}. The high temperature peak  is very sensitive to $J_2/|J_1|$,  which corresponds to the singlet-triplet excitation whose excitation energy is proportional to $J_2/|J_1|$, and in contrast, the low temperature peak slightly moves to the low temperature side with increasing $J_2/|J_1|$, which should be related to the single spin excitations. For $J_1>0$ and $J_2<0$, the specific heat exhibits a single peak behavior, which depends weakly on the rung coupling because the system is in the Haldane-like phase in this situation, as demonstrated in Fig.  \ref{capacity}(c). For the conventional square spin ladder, the specific heat differs from those of the HC ladder, where only single peak appears at low temperature for all three cases [Fig.  \ref{capacity}(d)], as there exist only singlet-triplet excitations at low temperature in this square ladder. At low temperature, by fitting our LTRG results we found that in the absence of a magnetic field the specific heat of the HC spin ladder at low temperature could have an asymptotic behavior of the form
\begin{equation}
C(T) \sim \frac{1}{T^\alpha} e^{-\Delta/T}, \label{C}
\end{equation}
where $\alpha$ is a constant. For all three cases in the HC ladder, we found $\alpha \approx 1$.  For the square spin ladder, our calculations give $\alpha \approx 3/2$ for the cases (i) and (ii), which is in agreement with the previous results \cite{Troyer}, while for case (iii) $\alpha \approx 1$.

\begin{figure}
  \includegraphics[width=1\linewidth]{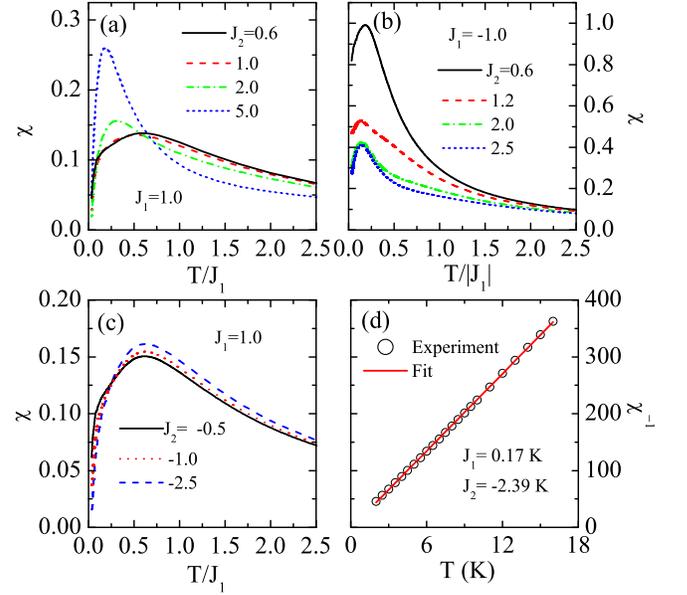}\\
  \caption{(Color online)  (a)-(c) The temperature dependence of susceptibility $\chi$ of the S=1/2 two-leg honeycomb spin ladders for different coupling ratios, which were obtained by the LTRG method with $D_c$=800. (d) Our LTRG result is fitted to the experimental data of the HC spin ladder compound $[Cu_2L^1(N_3)_4]_n$ (Ref. [\onlinecite{hcladder}]).}
  \label{susceptibility}
\end{figure}

The temperature dependence of the susceptibility ($\chi$) of the two-leg HC spin ladder for various cases is presented in Fig. \ref{susceptibility}. For all three cases under interest, $\chi$ reveals only one peak at low temperature and decays exponentially when $T \rightarrow 0$. However, for the two cases with $J_1>0$ a shoulder behavior is observed, while for $J_1<0$ there is no such feature. A closer inspection shows that for $J_1, J_2>0$ the shoulder and the broad peak merge into a sharp single peak when $J_2>2$ [Fig. \ref{susceptibility}(a)], and for $ J_2<0$ the shoulder becomes smeared with increasing $|J_2|$ [Fig. \ref{susceptibility}(c)]. For $J_1<0$ and $J_2>0$, the peak of $\chi$ is suppressed with increasing $J_2$ and remains almost intact, as indicated in Fig. \ref{susceptibility}(b). These behaviors of the susceptibility in the HC spin ladder are different from those of the corresponding square spin ladder. A fit to our numerical results shows that the low-temperature behavior of the susceptibility of the two-leg HC spin ladder could have the form of
\begin{equation}
\chi(T) \sim \frac{1}{T^\gamma} e^{-\Delta/T}, \label{C}
\end{equation}
where $\gamma$ is also a constant. For the cases (i) and (ii) in the HC ladder, we obtained $\gamma \approx 6/5$, and for case (iii)  $\gamma \approx 5/4$. For the square spin ladder, we found $\gamma \approx 1/2$ for the cases (i) and (ii), which recovers the previous results \cite{Troyer}, while for case (iii) $\gamma \approx 1$.

By fitting the experimental data of susceptibility of the HC spin ladder compound $[Cu_2L^1(N_3)_4]_n$ \cite{hcladder} with our LTRG result, we got the two coupling parameters $J_1=0.17 K$ and $J_2=-2.39 K$, as given by Fig. \ref{susceptibility}(d), suggesting that the rung coupling in this compound is ferromagnetic while the leg coupling is weakly antiferromagnetic, which is in good accordant with other experimental measurements \cite{hcladder}.

The Wilson ratio R$_w$, that is defined by
\begin{equation}\label{WR}
  R_w = \frac{4}{3} (\frac{\pi k_B}{g \mu_B})^{2} \frac{\chi T}{C},
\end{equation}
is an essential parameter to characterize the Fermi liquid \cite{WilR}. It is known that $R_w=1$ for noninteracting fermions because the specific heat $C$ depends linearly on temperature $T$ and the susceptibility $\chi$ is independent of $T$ at low temperatures. For the spin-1/2 single-impurity Kondo problem, it was shown that $R_w=2$ \cite{Stewart}. For the one-dimensional (1D) interacting electrons, the system demonstrates a  Tomonaga-Luttinger liquid (TLL) behavior, and the Wilson ratio has the form of $R_w=2/(1+v_s/v_c)$, which can be 1 in the noninteracting limit and approaches 2 in the strongly repulsive limit \cite{Schultz}, where $v_s$ and $v_c$ are the velocities of spin and charge excitations. For 1D antiferromagnets including spin-1/2 and spin-1 linear AF chains as well as the two-leg square spin ladder in magnetic fields that also show the TLL behavior in gapless regimes, the Wilson ratio bears  $R_w=4K_\sigma$ \cite{TLLW} with $K_\sigma$ the TLL parameter.

By means of the LTRG method with $D_c$=800, we calculated the Wilson ratio $R_w$ of the two-leg HC and square spin ladders at the lower critical magnetic field on which the gap corresponding to the $m=0$ plateau closes, respectively, for a comparison. The results are listed in Table \ref{wiltab}. It is seen that for both spin ladders the Wilson ratio has different values in three cases. For each case,  $R_w$'s of the square spin ladder are all larger than the corresponding $R_w$ values of the HC ladder. Note that our result of $R_w=3.98$ on the conventional square ladder with $J_1,J_2>0$ is consistent with that in Ref. [\onlinecite{TLLW}]. These results show that the low-energy physics of both spin ladders are indeed distinct.

\begin{table}[htbp]
\caption{Wilson ratio $R_{w}$ of the spin-1/2 two-leg square and honeycomb spin ladders at
the lower critical magnetic field for different types of couplings.}%
\begin{tabular}{m{1.5cm}m{1.5cm}m{2.0cm}m{2.0cm}}
\hline\hline
J$_{1}$ & J$_{2}$ & \multicolumn{2}{c}{%
\begin{tabular}{m{2.0cm}m{2.0cm}}
\multicolumn{2}{c}{R$_{w}$} \\ \hline
square & honeycomb %
\end{tabular}%
} \\ \hline
AF & AF & 3.98 & 1.21 \\
F & AF & 1.24 & 0.65 \\
AF & F & 4.57 & 1.75 \\ \hline\hline
\end{tabular}%
\label{wiltab}
\end{table}

 To summarize, we have systematically studied the unusual ground state and thermodynamic properties of spin-1/2 two-leg HC spin ladder by means of various methods. It is shown that the HC spin ladder has behaviors different remarkably from those of the conventional square ladder. A strong relevant term $\vec{n}\cdot\vec{J}$ is found in the HC ladder, which is hardly seen in other low-dimensional quantum spin systems. A half saturation magnetization plateau is observed, which is consistent with the condition for the occurrence of the topological quantization of magnetization and can be attributed to the formation of diluted dimer states, while it is absent in the square ladder. The ground state phase diagram of the HC spin ladder is presented for different cases, where the regions for the appearance of magnetization plateaux are carefully identified. The temperature dependences of the specific heat and susceptibility for various couplings are thoroughly explored in which two kinds of excitations are disclosed, and are also fairly compared with those of the square ladder. Our LTRG result is well fitted to the experimental result of a HC ladder compound. The Wilson ratios of the HC and square spin ladders after closure of energy gaps are found to be  quite distinct, manifesting that the square and HC ladders have different low-energy physics. We expect that the present study will stimulate more experimental and theoretical works on such fascinating two-leg HC spin ladder systems.

We are grateful to S. S. Gong, H. Krull, G. S. Uhrig, Y. J. Wang, Z. C. Wang, Q. R. Zheng and Z. G. Zhu for useful discussions. This work is supported in part by the NSFC (Grant No. 90922033 and No. 10934008), the MOST of China (Grant No. 2012CB932901, No. 2013CB933401), and the CAS.

\end{document}